\documentclass[tightenlines,floats,aps,nofootinbib,prd,twocolumn,preprintnumbers]{revtex4}

\usepackage[dvipdfmx]{graphicx}
\usepackage{amsmath,amssymb,algorithm,algorithmic,bm,color,mathrsfs}

\allowdisplaybreaks[1]

\newcommand{\kk}{\bm{k}}

\newcommand{\PP}{\mathcal{P}}
\newcommand{\calR}{\mathcal{R}}

\begin{document}

\title{Hunting for Statistical Anisotropy in Tensor Modes with B-mode Observations}

\author{Takashi Hiramatsu$^1$, Shuichiro Yokoyama$^{2,3}$, Tomohiro Fujita$^{4}$, Ippei Obata$^{5}$}

\affiliation{
$^1$ Department of Physics, Rikkyo University, Toshima, Tokyo, 171-8501, Japan\\
$^2$ Kobayashi Maskawa Institute,
Nagoya University,
Aichi 464-8602, Japan\\
$^3$ Kavli Institute for the Physics and Mathematics of the Universe
(Kavli IPMU, WPI), Todai Institutes for Advanced Study, the University of
Tokyo, Kashiwa 277-8583, Japan\\
$^4$ Department of Physics, Kyoto University, Kyoto 606-8502, Japan\\
$^5$ Institute for Cosmic Ray Research, The University of Tokyo,
5-1-5 Kashiwa-no-Ha, Kashiwa, Chiba, 277-8582, Japan
}

\begin{abstract}
We investigate a possibility of constraining statistical anisotropies of the primordial tensor perturbations
by using future observations for the Cosmic Microwave Background (CMB)
 B-mode polarization. By parameterizing a statistically-anisotropic
 tensor power spectrum as $P_h ({\bm k}) = P_h (k) \sum_n g_n \cos^n
 \theta_{\bm k}$, where $\theta_{\bm k}$ is an angle of the direction of
 $\hat{k}={\bm k}/k$ from a preferred direction, we find that it would be possible for future B-mode observations such as CMB-S4 to detect the tensor statistical anisotropy at the level of $g_n \sim {\mathcal O} (0.1)$.
\end{abstract}

\preprint{RUP-18-26}

\maketitle

%%%%%%%%%%%%%%%%%%%%%%%%%%%%%%%%%%%%%%%%%%%%%%%%%%%%%%%%%%%%%%%%%%%%%%%%%%%%%%%
%%%%%%%%%%%%%%%%%%%%%%%%%%%%%%%%%%%%%%%%%%%%%%%%%%%%%%%%%%%%%%%%%%%%%%%%%%%%%%%
\section{Introduction}
%%%%%%%%%%%%%%%%%%%%%%%%%%%%%%%%%%%%%%%%%%%%%%%%%%%%%%%%%%%%%%%%%%%%%%%%%%%%%%%
%%%%%%%%%%%%%%%%%%%%%%%%%%%%%%%%%%%%%%%%%%%%%%%%%%%%%%%%%%%%%%%%%%%%%%%%%%%%%%%

The detection of the B-mode polarization signal in Cosmic Microwave Background (CMB) 
is one of the most important challenges in cosmology, because it is sensitive to the primordial gravitational waves (PGWs) generated during inflation.
In standard inflationary scenario, the amplitude of the PGWs generated from vacuum fluctuations during inflation
is expected to depend only on the energy density of the inflation, $\rho_{\inf}$,
and hence its detection was considered as a direct probe for the scale of unknown physics.
The current constraint on the amplitude of the PGWs is  $r \lesssim 0.07$,
where $r$ represents the ratio of the power spectrum of the PGWs to that of the primordial curvature perturbations
 \cite{Ade:2015lrj, Array:2015xqh, Akrami:2018odb}. 
In 2020s, next generation CMB experiments such as LiteBIRD \cite{Matsumura:2013aja} and CMB-S4 \cite{Abazajian:2016yjj} are expected to achieve higher sensitivities reaching $r\simeq 10^{-3}$ which corresponds to 
$\rho_{\inf}^{1/4}\simeq 6\times 10^{15}$~GeV in the conventional vacuum fluctuation case.
Recently, other mechanisms of generating PGWs during inflation by introducing some matter fields (e.g. gauge fields) 
have been proposed,~\cite{Namba:2015gja, Dimastrogiovanni:2016fuu} and then this means that the vacuum PGWs are no longer the unique target of the B-mode observation.
Interestingly, the PGWs generated in these new mechanisms have not only the different relations between $r$ and $\rho_{\inf}$ but also observable signatures distinct from the vacuum one, 
{\it e.g.}, non-Gaussianity~\cite{Shiraishi:2016yun,Agrawal:2017awz}.
Among them, statistical anisotropies of the PGWs should be useful to distinguish the generation mechanisms and to extract richer information on the early universe from the B-mode observation.

The statistical anisotropy has been pursued mainly in the power spectrum of the curvature perturbation $P_\zeta$. This is because the anisotropic inflation and solid inflation models predict a quadrupole anisotropy in the curvature perturbation,
 $P_\zeta(\bm k) = P_{\zeta}(k)(1+g_*\cos^2\theta_{\bm k})$~\cite{Watanabe:2009ct, Watanabe:2010fh, Bartolo:2012sd, Fujita:2017lfu, Abolhasani:2013bpa,Bartolo:2013msa, Akhshik:2014gja, Bartolo:2014xfa}.
Furthermore, recent studies \cite{Kehagias:2017cym, Franciolini:2017ktv} argue that higher spin fields generate statistical anisotropies beyond quadrupole in $P_\zeta$ during inflation.
 Indeed, these kinds of anisotropies imprint interesting signatures in the CMB angular power spectrum.
 While in the standard picture CMB power spectra have only diagonal components in the angular multipole space due to a rotational invariance, statistical anisotropies can create specific nonzero off-diagonal correlations between temperature and polarization in CMB data.
 Several works have been discussed to test such kind of correlations due to the statistically-anisotropic curvature perturbation \cite{Watanabe:2010bu, Chen:2014eua, Bartolo:2014hwa, Bartolo:2017sbu, Franciolini:2018eno}.
So far, however, there is no evidence of the quadrupole anisotropy in $P_\zeta$ and we have an upper bound $|g_*| \lesssim 10^{-2}$ \cite{Kim:2013gka, Ade:2015lrj}.

 In this paper, we study the statistical anisotropy in the power spectrum of the PGWs.
Although little attention has been paid to the statistically-anisotropic PGWs, recent study \cite{Fujita:2018zbr} has proposed a model where large statistical anisotropies in $P_h$ can be generated.
In this model, U(1) gauge field is kinematically coupled to a spectator scalar field and gains a large background expectation value which breaks the isotropy of the universe, and then
due to the aniostropy of the universe
the perturbations of the spectator field and the gauge field could source the anisotropic tensor modes.
 Remarkably, the higher-order statistical anisotropies beyond quadrupole in $P_h$ can be predicted irrespective of the model parameters.
A similar prediction is also obtained when 2-form field takes over the role of the U(1) gauge field~\cite{Obata:2018ilf}.
 Other than this type of model, several works have suggested the generation of testable statistical anisotropies in tensor modes \cite{Shiraishi:2014owa, Chen:2014eua, Bartolo:2018igk}.
 Inspired by these predictions, we explore a possibility to test these higher statistical anisotropies of tensor modes through the B-mode angular power spectrum.
 We model the tensor statistical anisotropies as
 $P_h(\bm{k})=P_h(k)\sum_n g_n(k/k_0)^\gamma \cos^n\theta_{\bm k}$ and evaluate detectabilities of the coefficients $g_n$ in future missions.
 Compared with the previous study \cite{Shiraishi:2014owa}, we further investigate the sensitivities of $g_n$ up to $n=6$.

 This paper is organized as follows. In Sec.~\ref{sec:basic}, we describe basic equations for our Fisher analysis.
In Sec.~\ref{sec:results}, we obtain $1\sigma$ uncertainties of the anisotropic parameters, $g_n$ and $q_{LM}$.
We conclude in Sec.~\ref{sec:conclusion}.

%%%%%%%%%%%%%%%%%%%%%%%%%%%%%%%%%%%%%%%%%%%%%%%%%%%%%%%%%%%%%%%%%%%%%%%%%%%%%%%
%%%%%%%%%%%%%%%%%%%%%%%%%%%%%%%%%%%%%%%%%%%%%%%%%%%%%%%%%%%%%%%%%%%%%%%%%%%%%%%
\section{Basic equations}
\label{sec:basic}
%%%%%%%%%%%%%%%%%%%%%%%%%%%%%%%%%%%%%%%%%%%%%%%%%%%%%%%%%%%%%%%%%%%%%%%%%%%%%%%
%%%%%%%%%%%%%%%%%%%%%%%%%%%%%%%%%%%%%%%%%%%%%%%%%%%%%%%%%%%%%%%%%%%%%%%%%%%%%%%

%%==============================================================
%%==============================================================
\subsection{Anisotropies}
%%==============================================================
%%==============================================================

Harmonic coefficients of B-mode anisotropies induced by the
primordial tensor perturbations $h_{\pm 2}(\kk)$ can be written 
in terms of the transfer function
$T_\ell^{(B)}(k)$ (see {\it e.g.} Ref.~\cite{Bartolo:2014hwa}),
%%%%%%%%%%%%%%%%%%%%%%%%%%%%%%%%%%%%%%%%%%%%%%%%%%%%%%%%%%%%%%%%%%%%
%
\begin{align}
a^{(B)}_{\ell m}
 = 4\pi(-i)^\ell\int\frac{d^3k}{(2\pi)^3}\sum_{s=\pm 2} h_s(\kk)
   T_\ell^{(B)}(k){}_{-s}Y^*_{\ell m}(\hat{k}),
\end{align}
%
%%%%%%%%%%%%%%%%%%%%%%%%%%%%%%%%%%%%%%%%%%%%%%%%%%%%%%%%%%%%%%%%%%%%
with ${}_{s}Y_{\ell m}$ being the spin-$s$ spherical harmonics.
The power spectrum of the tensor perturbations is defined as
%%%%%%%%%%%%%%%%%%%%%%%%%%%%%%%%%%%%%%%%%%%%%%%%%%%%%%%%%%%%%%%%%%%%
%
\begin{align}
\langle h_{+2}(\kk_1)h_{-2}(\kk_2) \rangle
 = \frac{1}{2}(2\pi)^3P_{h}(\kk_1)\delta^{(3)}(\kk_1-\kk_2),
\end{align}
%
%%%%%%%%%%%%%%%%%%%%%%%%%%%%%%%%%%%%%%%%%%%%%%%%%%%%%%%%%%%%%%%%%%%%
where we have used $h_{-2}({\bm k}) = h^\ast_{+2}({\bm k})$.
If the rotational invariance is broken, the power spectrum
could have the directional dependence, which can be parameterized as \cite{Valentini:2015sna},
%%%%%%%%%%%%%%%%%%%%%%%%%%%%%%%%%%%%%%%%%%%%%%%%%%%%%%%%%%%%%%%%%%%%
%
\begin{align}
P_{h}(\kk) = P_{h}(k)\sum_{LM}Q_{LM}(k)Y_{LM}(\hat{k}),
\label{eq:aniso1}
\end{align}
%
%%%%%%%%%%%%%%%%%%%%%%%%%%%%%%%%%%%%%%%%%%%%%%%%%%%%%%%%%%%%%%%%%%%%
with $L$ running over even numbers, $0,2,4,.\ldots$, where $P_h(k)$ is the isotropic (monopole) part,
and $\hat{k}:=\kk/k$.
Taking into account the directional dependence, we obtain
the correlation of the harmonic coefficients \cite{Shiraishi:2014owa},
%%%%%%%%%%%%%%%%%%%%%%%%%%%%%%%%%%%%%%%%%%%%%%%%%%%%%%%%%%%%%%%%%%%%
%
\begin{align}
C^{BB}_{\ell_1m_1;\ell_2m_2}
&:=\langle a^{(B)}_{\ell_1 m_1} a^{(B)*}_{\ell_2 m_2} \rangle
\notag\\
 &= \frac{2}{\pi}i^{\ell_2-\ell_1}
 (-1)^{m_1}
\delta^{\rm even}_{\ell_1+\ell_2}
\sum_{LM}
 \mathcal{G}^{-m_1m_2M;-220}_{\ell_1\ell_2L}
\notag \\ & \quad 
\times  \int\!dk\,k^2
  P_{h}(k)
Q_{LM}(k)
 T^{(B)}_{\ell_1}(k)
 T^{(B)}_{\ell_2}(k),
\label{eq:Clmlm1}
\end{align}
%
%%%%%%%%%%%%%%%%%%%%%%%%%%%%%%%%%%%%%%%%%%%%%%%%%%%%%%%%%%%%%%%%%%%%
where 
$\delta^{\rm even}_{a}$ is 1 if $a$ is even, and 0 otherwise, and 
$\mathcal{G}^{m_1m_2m_3;s_1s_2s_3}_{\ell_1\ell_2\ell_3}$ is the spin-weighted Gaunt integral 
that is written in terms of the product of Wigner's $3j$-symbols,
%%%%%%%%%%%%%%%%%%%%%%%%%%%%%%%%%%%%%%%%%%%%%%%%%%%%%%%%%%%%%%%%%%%%
%
\begin{align}
\mathcal{G}^{m_1m_2m_3;s_1s_2s_3}_{\ell_1\ell_2\ell_3}
&:=
\int\!d\Omega\,{}_{s_1}Y_{\ell_1m_1}(\Omega){}_{s_2}Y_{\ell_2m_2}(\Omega){}_{s_3}Y_{\ell_3m_3}(\Omega)\notag \\
&=\sqrt{\frac{(2\ell_1+1)(2\ell_2+1)(2\ell_3+1)}{4\pi}}
\notag \\ &\times 
\begin{pmatrix}
  \ell_1 & \ell_2 & \ell_3 \\
  m_1 & m_2 & m_3
\end{pmatrix}
\begin{pmatrix}
  \ell_1 & \ell_2 & \ell_3 \\
  -s_1 & -s_2 & -s_3
\end{pmatrix}.
\end{align}
%
%%%%%%%%%%%%%%%%%%%%%%%%%%%%%%%%%%%%%%%%%%%%%%%%%%%%%%%%%%%%%%%%%%%%
In the present study, we assume that the scale dependence of the
anisotropic parameter is given as \cite{Shiraishi:2014owa}
%%%%%%%%%%%%%%%%%%%%%%%%%%%%%%%%%%%%%%%%%%%%%%%%%%%%%%%%%%%%%%%%%%%%
%
\begin{align}
 Q_{LM}(k) = q_{LM}\left(\frac{k}{k_0}\right)^\gamma,
 \label{eq:sdep}
\end{align}
%
%%%%%%%%%%%%%%%%%%%%%%%%%%%%%%%%%%%%%%%%%%%%%%%%%%%%%%%%%%%%%%%%%%%%
with constants $q_{LM}$ and $\gamma$. Finally, Eq.~(\ref{eq:Clmlm1}) reads
%%%%%%%%%%%%%%%%%%%%%%%%%%%%%%%%%%%%%%%%%%%%%%%%%%%%%%%%%%%%%%%%%%%%
%
\begin{align}
C^{BB}_{\ell_1m_1;\ell_2m_2}(\gamma)
 &= \frac{2}{\pi}i^{\ell_2-\ell_1}
 (-1)^{m_1}
\notag \\ & \quad 
\times  
\sum_{LM}
\delta^{\rm even}_{\ell_1+\ell_2+L}
 \mathcal{G}^{-m_1m_2M;-220}_{\ell_1\ell_2L}
q_{LM}C^{BB}_{\ell_1\ell_2}(\gamma),
\end{align}
%
%%%%%%%%%%%%%%%%%%%%%%%%%%%%%%%%%%%%%%%%%%%%%%%%%%%%%%%%%%%%%%%%%%%%
where
%%%%%%%%%%%%%%%%%%%%%%%%%%%%%%%%%%%%%%%%%%%%%%%%%%%%%%%%%%%%%%%%%%%%
%
\begin{align}
 C^{BB}_{\ell_1\ell_2}(\gamma) :=
\frac{2}{\pi}
  \int\!dk\,k^2
P_{h}(k)
 T^{(B)}_{\ell_1}(k)
 T^{(B)}_{\ell_2}(k)
 \left(\frac{k}{k_0}\right)^\gamma.
 \label{eq:Cl1l2}
\end{align}
%
%%%%%%%%%%%%%%%%%%%%%%%%%%%%%%%%%%%%%%%%%%%%%%%%%%%%%%%%%%%%%%%%%%%%
Note that for the case with $q_{LM}=\delta_{L0}\delta_{M0}$, that is, statistically-isotropic power spectrum, one can find that
the above expression is equivalent to the standard form of the angular power spectrum.

In the theoretical models which predict the statistical anisotropy in primordial tensor modes,
the statistical anisotropy is often parameterized
in terms of the power series of the cosine function (e.g., \cite{Kim:2013gka, Ade:2015lrj}), as
%%%%%%%%%%%%%%%%%%%%%%%%%%%%%%%%%%%%%%%%%%%%%%%%%%%%%%%%%%%%%%%%%%%%
%
\begin{align}
 P_{h}(\kk) &= P_{h}(k)\sum^N_{n={\rm even}}g_n\left(\frac{k}{k_0}\right)^\gamma\cos^n \theta_{\bm k},
\label{eq:aniso2}
\end{align}
%
%%%%%%%%%%%%%%%%%%%%%%%%%%%%%%%%%%%%%%%%%%%%%%%%%%%%%%%%%%%%%%%%%%%%
where $\theta_{\bm k}$ measures the angle of the direction of $\hat{k}$ from
a preferred direction. 
Thus, it should be useful to give a relation between the parameters $g_n$ and $q_{LM}$, and
according to Eq.~(\ref{eq:qg})
their relations are given by
%%%%%%%%%%%%%%%%%%%%%%%%%%%%%%%%%%%%%%%%%%%%%%%%%%%%%%%%%%%%%%%%%%%%
%
\begin{align}
q_{0M} &= 2\sqrt{\pi}\left(g_0+\frac{g_2}{3}+\frac{g_4}{5}+\frac{g_6}{7}\right)\delta_{M0},\\
q_{2M} &= 4\sqrt{\frac{\pi}{5}}\left(\frac{g_2}{3}+\frac{2}{7}g_4+\frac{5}{21}g_6\right)\delta_{M0},\\
q_{4M} &= 16\sqrt{\pi}\left(\frac{g_4}{105}+\frac{g_6}{77}\right)\delta_{M0},\\
q_{6M} &= \frac{32}{231}\sqrt{\frac{\pi}{13}}\,g_6\,\delta_{M0}.
\end{align}
%
%%%%%%%%%%%%%%%%%%%%%%%%%%%%%%%%%%%%%%%%%%%%%%%%%%%%%%%%%%%%%%%%%%%%

%%==============================================================
%%==============================================================
\subsection{Fisher information matrix}
%%==============================================================
%%==============================================================

To quantify the $1\sigma$ uncertainties of the anisotropic parameters, $\{q_{LM}\}$
or $\{g_n\}$, we use the Fisher information matrix.
The details of the computation of the Fisher information matrix in our
study are provided
in Appendix \ref{app:fisher}. Here we consider only the B-mode in the full
expression in Eq.~(\ref{eq:Fisher_aniso_q}) or Eq.~(\ref{eq:Fisher_aniso_g})
with Eq.~(\ref{eq:Fisher_kernel}), and it reads
%%%%%%%%%%%%%%%%%%%%%%%%%%%%%%%%%%%%%%%%%%%%%%%%%%%%%%%%%%%%%%%%%%%%
%
\begin{align}
 F^{BB}_L
&=\frac{f_{\rm sky}}{4\pi}  \sum_{\ell_1\ell_2}(2\ell_1+1)(2\ell_2+1)
\begin{pmatrix}
 \ell_1 & \ell_2 & L \\
 -2 & 2 & 0
\end{pmatrix}^2
\notag \\ & \quad \times
\frac{\left(C^{BB}_{\ell_1\ell_2}\right)^2}{\widetilde{C}^{BB}_{\ell_1}\widetilde{C}^{BB}_{\ell_2}},
\label{eq:Fisher_kernel_BB}
\end{align}
%
%%%%%%%%%%%%%%%%%%%%%%%%%%%%%%%%%%%%%%%%%%%%%%%%%%%%%%%%%%%%%%%%%%%%
where $\widetilde{C}^{BB}_{\ell}$ is the total angular power spectrum 
of B-mode polarization defined in Eq.~(\ref{eq:totalCl}).
Using Eq.~(\ref{eq:Fisher_aniso_q}) or Eq.~(\ref{eq:Fisher_aniso_g}), 
we can estimate the uncertainties of the measurement of the anisotropic parameters,
%%%%%%%%%%%%%%%%%%%%%%%%%%%%%%%%%%%%%%%%%%%%%%%%%%%%%%%%%%%%%%%%%%%%
%
\begin{align}
 \sigma_{q_{LM}}^2 = (F_{LM;LM})^{-1},
\quad
 \sigma_{g_n}^2 = (F_{nn})^{-1}.
\label{eq:def_error}
\end{align}
%
%%%%%%%%%%%%%%%%%%%%%%%%%%%%%%%%%%%%%%%%%%%%%%%%%%%%%%%%%%%%%%%%%%%%
A noise model we adopt in the present study is
%%%%%%%%%%%%%%%%%%%%%%%%%%%%%%%%%%%%%%%%%%%%%%%%%%%%%%%%%%%%%%%%%%%%
%
\begin{align}
 \mathcal{N}^{BB}_\ell &= N^{BB}_\ell e^{\ell^2\sigma_b^2}, 
\end{align}
%
%%%%%%%%%%%%%%%%%%%%%%%%%%%%%%%%%%%%%%%%%%%%%%%%%%%%%%%%%%%%%%%%%%%%
in which we assume the detector noise
$N^{BB}_\ell$ and the beam effect $\sigma_b$ are parameterized as
\cite{Katayama:2011eh}
%%%%%%%%%%%%%%%%%%%%%%%%%%%%%%%%%%%%%%%%%%%%%%%%%%%%%%%%%%%%%%%%%%%%
%
\begin{align}
 N^{BB}_\ell &= \left(\frac{\pi}{10800}\frac{w_{BB}{}^{-1/2}}{\mu{\rm
           K\;arcmin}}\right)^2\mu{\rm K}^2\;{\rm str},
\label{eq:def_NBB}
\\
\sigma_b&=\frac{\pi}{10800}\frac{\theta_{\rm FWHM}}{\rm arcmin}\frac{1}{\sqrt{8\ln 2}}.
\end{align}
%
%%%%%%%%%%%%%%%%%%%%%%%%%%%%%%%%%%%%%%%%%%%%%%%%%%%%%%%%%%%%%%%%%%%%
with $\theta_{\rm FWHM}$ being the full width at half maximum (FWHM) of the beam 
in the unit of arcmin. 
Although we do not take into account neither
the lensing effect from the E-mode induced 
by scalar perturbations nor the foreground noises sourced by dust emission,
it is possible to emulate the cases including them by increasing
the noise parameter $w_{BB}^{-1/2}$.

%%==============================================================
%%==============================================================
\section{Detectability of statistical anisotropies of tensor perturbations}
\label{sec:results}
%%==============================================================
%%==============================================================

We use {\tt cmb2nd}\footnote{This Boltzmann code is not public yet, but
we have confirmed that the transfer functions obtained from this
precisely agree with those from CAMB ({\tt https://camb.info/}). See also
Ref.~\cite{Hiramatsu:2018nfa} in which we used the same code.} to compute the transfer function of B-mode with the
cosmological parameters from the Planck 2015 results
(TT,TE,EE+lowP+lensing+ext in Ref.~\cite{Ade:2015xua}), which are tabulated in
Table \ref{tab:param}. 
We assume a $0.5$ degree FWHM beam
(designed in LiteBIRD \cite{Matsumura:2013aja}) and the noise 
level with  $w_{\rm BB}^{-1/2}=1.0, 5.0, 63.1\mu {\rm K}\cdot {\rm arcmin}$
which correspond to CMB-S4~\cite{Abazajian:2016yjj}, LiteBIRD~\cite{Matsumura:2013aja}, and Planck~\cite{Planck:2006aa},
respectively.
In addition to them, we also compute the cosmic-variance-limited (CVL) case
with $w_{BB}^{-1/2}=0$.

\begin{table}[ht]
 \begin{tabular}{lll}
\hline
\hline
\multicolumn{2}{c}{parameter} & \multicolumn{1}{c}{value} \\
\hline
amplitude of curvature perturbation & $\PP_{\calR 0}$ & $2.384\times 10^{-9}$ \\
tensor-to-scalar ratio        & $r$       & 0.01 \\
pivot scale                         & $k_{\rm pivot}$ & $0.002~{\rm Mpc}^{-1}$ \\
spectral index                      & $n_s$           & $0.9667$ \\
reduced Hubble parameter            & $h$             & $0.6774$ \\
dark matter fraction                & $h^2\Omega_{\rm CDM}$ & $0.1188$ \\
baryon fraction                     & $h^2\Omega_{\rm b}$   & $0.02230$ \\
effective number of neutrinos       & $N_{\rm eff}$         & $3.046$ \\
photon's temperature                & $T_{\gamma,0}$        & $2.7255~{\rm K}$ \\
optical depth                       & $\tau$                & 0.066 \\
Helium abundance                    & $Y_p$                 & 0.24667 \\
\hline
\hline
 \end{tabular}
  \caption{Cosmological parameters used in the present study. The
 all parameters except for the tensor-to-scalar ratio are
 provided by Planck 2015 results (TT,TE,EE+lowP+lensing+ext in
 Ref.~\cite{Ade:2015xua}). The amplitude of curvature perturbation and
 the tensor-to-scalar ratio are evaluated at $k=k_{\rm pivot}$, and we
 assume $n_{s,0.002}=n_{s,0.05}$ in the notation of Ref.~\cite{Ade:2015xua}.} 
  \label{tab:param}
\end{table}

\begin{table}[!ht]
 \begin{tabular}{|c||cccc|}
\hline 
      & CVL  & 1.0   & 5.0   & 63.1\\ 
\hline % m359[a-d]
$g_{2}$ & $1.93\times 10^{-3}$ & $5.98\times 10^{-2}$ & $2.03\times
              10^{-1}$ & $3.25$ \\
\hline
\hline % m360[a-d]
$g_{2}$ & $8.23\times 10^{-3}$ & $2.66\times 10^{-1}$ & $9.25\times
              10^{-1}$ & $1.89\times 10^{1}$ \\
$g_{4}$ & $1.24\times 10^{-2}$ & $3.90\times 10^{-1}$ & $1.35$ &
                  $2.94\times 10^{1}$ \\
\hline
\hline % m361[a-d]
$g_{2}$ & $2.33\times 10^{-2}$ & $6.29\times 10^{-1}$ & $1.47$ &
                  $1.92\times 10^{1}$ \\
$g_{4}$ & $8.27\times 10^{-2}$ & $2.25$ & $4.67$ & $3.23\times
                  10^{1}$ \\
$g_{6}$ & $6.46\times 10^{-2}$ & $1.80$ & $3.67$ & $1.11\times
                  10^{1}$ \\
\hline
\hline % m361[a-d]
$q_{2M}$ & $4.98\times 10^{-3}$ & $1.26\times 10^{-1}$ & $4.10\times
              10^{-1}$ & $9.52$ \\
$q_{4M}$ & $4.28\times 10^{-3}$ & $1.53\times 10^{-1}$ & $5.59\times
              10^{-1}$ & $1.27\times 10^{1}$ \\
$q_{6M}$ & $5.71\times 10^{-3}$ & $1.50\times 10^{-1}$ & $2.66\times
              10^{-1}$ & $7.55\times 10^{-1}$ \\
\hline
 \end{tabular}
  \caption{$\sigma_{g_n}$ for $w_{BB}^{-1/2}=63.1, 5.0,
 1.0\mu$K$\cdot$arcmin and the CVL case with $\gamma=0$ and $f_{\rm sky}=1$.}
  \label{fig:resultBB_g0}
\end{table}

 The $1\sigma$ uncertainties of the measurements of $g_n$ and $q_{LM}$ 
with even numbers of $n$ and $L$
up to $n,L\leq 6$ are summarized in 
Tables ~\ref{fig:resultBB_g0}-\ref{fig:resultBB_gp}
which are computed with fixed $\gamma$; $\gamma=0$
(Table \ref{fig:resultBB_g0}), $\gamma=-1,-1/2$ (Table \ref{fig:resultBB_gm})
and $\gamma=1/2,1$ (Table \ref{fig:resultBB_gp}.)
Throughout this paper, the isotropic part of angular power spectrum
is supposed to be $C^{BB}_{\ell} := C^{BB}_{\ell\ell}(\gamma=0)$ with $r=0.01$ at $k=k_{\rm pivot}$.
Hence the observed signal is given as
%%%%%%%%%%%%%%%%%%%%%%%%%%%%%%%%%%%%%%%%%%%%%%%%%%%%%%%%%%%%%%%%%%%%
%
\begin{align}
C^{\rm obs}_{\ell_1m_1;\ell_2m_2}(\gamma)=C^{BB}_{\ell_1}\delta_{\ell_1\ell_2}\delta_{m_1m_2} + C^{BB}_{\ell_1m_1;\ell_2m_2}(\gamma).
\end{align}
%
%%%%%%%%%%%%%%%%%%%%%%%%%%%%%%%%%%%%%%%%%%%%%%%%%%%%%%%%%%%%%%%%%%%%
In the case with $\gamma=0$ (Table \ref{fig:resultBB_g0}), we fix
$g_0=1$ or $q_{00}=1$ and vary $g_n$ ($q_{LM}$) for $n \geq 2$
($L\geq 2$), whereas in the cases with $\gamma\ne0$ (Table
\ref{fig:resultBB_gm} and \ref{fig:resultBB_gp}), we vary also $g_0$ or
$q_{00}$.
In the tables, we show the results with $f_{\rm sky}=1$. One can obtain
those with $f_{\rm sky}<1$ by multiplying the values by $1/\sqrt{f_{\rm sky}}$.

Note that $\sigma_{g_n}$ with $n\leq N$ has a strong dependence on the
number of parameters $N$ due to the non-vanishing off-diagonal
components of the Fisher information matrix, whereas $\sigma_{q_{LM}}$
is independent of $N$ since the corresponding Fisher information matrix is
diagonal. Hence, as for $\sigma_{g_n}$, we compute their uncertainties
for $N=2,4,6$, respectively. 

In Table \ref{fig:resultBB_g0}, we find that,
when we take into account both $g_4$ and $g_6$, their uncertainties are greater than the unity
even in the case with $w_{BB}^{-1/2}=1~\mu K\cdot$arcmin, which leads to the difficulty 
of measurement of such higher-order anisotropies. On the other hand, when we take into account
up to $g_4$, it is implied that 
the hexadecapole anisotropy with $g_4=\mathcal{O}(1)$ can be detected by an observatory
whose specification is similar to CMB-S4.

In Table \ref{fig:resultBB_gm} and \ref{fig:resultBB_gp}, we estimate
the uncertainties with various $\gamma$. 
In the red-tilted
cases, our results even with $\gamma=-1$ indicate the possibility to detect
$g_2$ by LiteBIRD and CMB-S4, while it is fairly difficult to get a
signal of $g_4$ even with CMB-S4. On the other hand, in the blue-tilted
cases, we can marginally detect $g_4$, since much power is induced to the
angular power spectrum on large $\ell$.
Note that, in Ref.~\cite{Shiraishi:2014owa}, the authors reported the 
uncertainties with $\gamma=-2$, $\sigma_{q_{0M}}=30$ and
$\sigma_{q_{2M}}=58$ in our notations. In the present study, we 
obtained $\sigma_{g_0}=20.6$ and $\sigma_{g_2}=56.4$ with
$\gamma=-2$ in the
CVL case, which are well consistent to the previous results.

\onecolumngrid

\begin{table}[!ht]
 \begin{tabular}{|c||cccc|}
\hline
      & CVL  & 1.0   & 5.0   & 63.1\\ \hline
\hline % m371[a-d]
$g_{0}$ & $5.28\times 10^{-2}$ & $9.67\times 10^{-2}$ & $1.31\times
              10^{-1}$ & $5.93\times 10^{-1}$ \\
$g_{2}$ & $1.43\times 10^{-1}$ & $2.61\times 10^{-1}$ & $3.58\times
              10^{-1}$ & $1.64$ \\
\hline
\hline % m372[a-d]
$g_{0}$ & $7.23\times 10^{-2}$ & $1.56\times 10^{-1}$ & $2.23\times
              10^{-1}$ & $1.04$ \\
$g_{2}$ & $5.14\times 10^{-1}$ & $1.25$ & $1.85$ & $8.65$ \\
$g_{4}$ & $5.76\times 10^{-1}$ & $1.42$ & $2.11$ & $9.91$ \\
\hline
\hline % m379[a-d]
$g_{0}$ & $7.38\times 10^{-2}$ & $1.56\times 10^{-1}$ & $2.24\times
              10^{-1}$ & $1.04$ \\
$g_{2}$ & $6.02\times 10^{-1}$ & $1.29$ & $1.88$ & $8.69$ \\
$g_{4}$ & $1.10$ & $1.72$ & $2.33$ & $1.02\times 10^{1}$ \\
$g_{6}$ & $6.91\times 10^{-1}$ & $7.08\times 10^{-1}$ & $7.17\times
              10^{-1}$ & $1.88$ \\
\hline
\hline % m379[a-d]
$q_{0M}$ & $8.08\times 10^{-2}$ & $1.49\times 10^{-1}$ & $1.88\times
              10^{-1}$ & $8.21\times 10^{-1}$ \\
$q_{2M}$ & $1.51\times 10^{-1}$ & $2.76\times 10^{-1}$ & $3.79\times
              10^{-1}$ & $1.73$ \\
$q_{4M}$ & $1.56\times 10^{-1}$ & $3.84\times 10^{-1}$ & $5.71\times
              10^{-1}$ & $2.68$ \\
$q_{6M}$ & $4.70\times 10^{-2}$ & $4.82\times 10^{-2}$ & $4.89\times
              10^{-2}$ & $1.28\times 10^{-1}$ \\
\hline
 \end{tabular}
 \begin{tabular}{|c||cccc|}
\hline
      & CVL  & 1.0   & 5.0   & 63.1\\ \hline
\hline % m373[a-d]
$g_{0}$ & $1.23\times 10^{-2}$ & $8.26\times 10^{-2}$ & $1.96\times
              10^{-1}$ & $1.42$ \\
$g_{2}$ & $3.36\times 10^{-2}$ & $2.18\times 10^{-1}$ & $5.26\times
              10^{-1}$ & $3.96$ \\
\hline
\hline % m374[a-d]
$g_{0}$ & $1.59\times 10^{-2}$ & $1.22\times 10^{-1}$ & $3.23\times
              10^{-1}$ & $2.38$ \\
$g_{2}$ & $1.06\times 10^{-1}$ & $9.23\times 10^{-1}$ & $2.62$ &
                  $1.94\times 10^{1}$ \\
$g_{4}$ & $1.17\times 10^{-1}$ & $1.05$ & $2.99$ & $2.22\times
                  10^{1}$ \\
\hline
\hline % m380[a-d]
$g_{0}$ & $2.02\times 10^{-2}$ & $1.27\times 10^{-1}$ & $3.25\times
              10^{-1}$ & $2.38$ \\
$g_{2}$ & $2.82\times 10^{-1}$ & $1.19$ & $2.74$ & $1.95\times
                  10^{1}$ \\
$g_{4}$ & $7.93\times 10^{-1}$ & $2.49$ & $3.84$ & $2.31\times
                  10^{1}$ \\
$g_{6}$ & $5.75\times 10^{-1}$ & $1.66$ & $1.77$ & $4.66$ \\
\hline
\hline % m380[a-d]
$q_{0M}$ & $1.82\times 10^{-2}$ & $1.39\times 10^{-1}$ & $3.07\times
              10^{-1}$ & $1.90$ \\
$q_{2M}$ & $3.55\times 10^{-2}$ & $2.30\times 10^{-1}$ & $5.56\times
              10^{-1}$ & $4.18$ \\
$q_{4M}$ & $3.17\times 10^{-2}$ & $2.83\times 10^{-1}$ & $8.09\times
              10^{-1}$ & $5.99$ \\
$q_{6M}$ & $3.92\times 10^{-2}$ & $1.13\times 10^{-1}$ & $1.20\times
              10^{-1}$ & $3.18\times 10^{-1}$ \\
\hline
 \end{tabular}
  \caption{$\sigma_{g_n}$ for $w_{BB}^{-1/2}=63.1, 5.0,
 1.0\mu$K$\cdot$arcmin and the CVL case with
 $\gamma=-1$ (left), $-1/2$ (right) and $k_0=k_{\rm pivot}$ and $f_{\rm sky}=1$.}
  \label{fig:resultBB_gm}
\end{table}

\begin{table}[ht]
 \begin{tabular}{|c||cccc|}
\hline
      & CVL  & 1.0   & 5.0   & 63.1\\ \hline
$g_{0}$ & $2.00\times 10^{-4}$ & $2.31\times 10^{-2}$ & $8.37\times
              10^{-2}$ & $5.37$ \\
$g_{2}$ & $5.47\times 10^{-4}$ & $5.94\times 10^{-2}$ & $2.08\times
              10^{-1}$ & $1.48\times 10^{1}$ \\
\hline
\hline % m376[a-d]
$g_{0}$ & $2.55\times 10^{-4}$ & $3.37\times 10^{-2}$ & $1.25\times
              10^{-1}$ & $8.29$ \\
$g_{2}$ & $1.67\times 10^{-3}$ & $2.53\times 10^{-1}$ & $9.56\times
              10^{-1}$ & $6.49\times 10^{1}$ \\
$g_{4}$ & $1.84\times 10^{-3}$ & $2.86\times 10^{-1}$ & $1.09$ &
                  $7.37\times 10^{1}$ \\
\hline
\hline % m381[a-d]
$g_{0}$ & $3.29\times 10^{-4}$ & $4.36\times 10^{-2}$ & $1.61\times
              10^{-1}$ & $8.31$ \\
$g_{2}$ & $4.67\times 10^{-3}$ & $6.34\times 10^{-1}$ & $2.32$ &
                  $6.59\times 10^{1}$ \\
$g_{4}$ & $1.32\times 10^{-2}$ & $1.77$ & $6.44$ & $8.12\times
                  10^{1}$ \\
$g_{6}$ & $9.59\times 10^{-3}$ & $1.28$ & $4.65$ & $2.48\times
                  10^{1}$ \\
\hline
\hline % m381[a-d]
$q_{0M}$ & $2.92\times 10^{-4}$ & $4.20\times 10^{-2}$ & $1.67\times
              10^{-1}$ & $7.49$ \\
$q_{2M}$ & $5.78\times 10^{-4}$ & $6.28\times 10^{-2}$ & $2.19\times
              10^{-1}$ & $1.57\times 10^{1}$ \\
$q_{4M}$ & $4.97\times 10^{-4}$ & $7.74\times 10^{-2}$ & $2.94\times
              10^{-1}$ & $1.99\times 10^{1}$ \\
$q_{6M}$ & $6.53\times 10^{-4}$ & $8.71\times 10^{-2}$ & $3.17\times
              10^{-1}$ & $1.69$ \\
\hline
 \end{tabular}
 \begin{tabular}{|c||cccc|}
\hline
      & CVL  & 1.0   & 5.0   & 63.1\\ \hline
\hline % m377[a-d]
$g_{0}$ & $2.11\times 10^{-5}$ & $1.10\times 10^{-2}$ & $4.19\times
              10^{-2}$ & $4.01$ \\
$g_{2}$ & $5.76\times 10^{-5}$ & $2.80\times 10^{-2}$ & $1.02\times
              10^{-1}$ & $1.02\times 10^{1}$ \\
\hline
\hline % m378[a-d]
$g_{0}$ & $2.72\times 10^{-5}$ & $1.61\times 10^{-2}$ & $6.18\times
              10^{-2}$ & $5.95$ \\
$g_{2}$ & $1.80\times 10^{-4}$ & $1.22\times 10^{-1}$ & $4.66\times
              10^{-1}$ & $4.51\times 10^{1}$ \\
$g_{4}$ & $1.99\times 10^{-4}$ & $1.38\times 10^{-1}$ & $5.30\times
              10^{-1}$ & $5.12\times 10^{1}$ \\
\hline
\hline % m382[a-d]
$g_{0}$ & $3.47\times 10^{-5}$ & $2.10\times 10^{-2}$ & $8.33\times
              10^{-2}$ & $6.04$ \\
$g_{2}$ & $4.88\times 10^{-4}$ & $3.06\times 10^{-1}$ & $1.26$ &
                  $5.02\times 10^{1}$ \\
$g_{4}$ & $1.38\times 10^{-3}$ & $8.53\times 10^{-1}$ & $3.56$ &
                  $8.40\times 10^{1}$ \\
$g_{6}$ & $9.98\times 10^{-4}$ & $6.17\times 10^{-1}$ & $2.58$ &
                  $4.88\times 10^{1}$ \\
\hline
\hline % m382[a-d]
$q_{0M}$ & $3.13\times 10^{-5}$ & $2.05\times 10^{-2}$ & $8.61\times
              10^{-2}$ & $7.59$ \\
$q_{2M}$ & $6.08\times 10^{-5}$ & $2.96\times 10^{-2}$ & $1.08\times
              10^{-1}$ & $1.08\times 10^{1}$ \\
$q_{4M}$ & $5.38\times 10^{-5}$ & $3.73\times 10^{-2}$ & $1.43\times
              10^{-1}$ & $1.38\times 10^{1}$ \\
$q_{6M}$ & $6.80\times 10^{-5}$ & $4.20\times 10^{-2}$ & $1.76\times
              10^{-1}$ & $3.32$ \\
\hline
 \end{tabular}
  \caption{$\sigma_{g_n}$ for $w_{BB}^{-1/2}=63.1, 5.0,
 1.0\mu$K$\cdot$arcmin and the CVL case with
 $\gamma=1/2$ (left), 1 (right) and $k_0=k_{\rm pivot}$ and $f_{\rm sky}=1$.}
  \label{fig:resultBB_gp}
\end{table}

\twocolumngrid

Let us compare our result with theoretical predictions. 
The model in Ref.~\cite{Fujita:2018zbr} predicts $g_0=1, g_2=-1, g_4=1, g_6=-1$ irrespective of the model
parameters, while $\gamma\lesssim-1/2$ is required to produce a detectable
amplitude of the sourced PGW (i.e. $r_{\rm source}\gtrsim 10^{-3}$). In
the case of $\gamma=-1, -1/2$ (Table \ref{fig:resultBB_gm}), these result
show that the predicted $g_0$ and $g_2$ are marginally detectable,
whereas it is challenging to measure $g_4$ and $g_6$ even with the
CMB-S4 experiment at $1\sigma$ level. 
Let us also consider the discrimination between the models.
The prediction in Ref.~\cite{Obata:2018ilf} is
$g_0=1, g_2=1, g_4=-2, g_6=1$, and the sign of $g_2$ is flipped from that of Ref.~\cite{Fujita:2018zbr}.
This difference  is originated in the distinction between the particle types which generate the PGWs (i.e. U(1) gauge field or 2-form field). 
Therefore, once $g_n~(n\ge2)$ is detected, we may gain an insight what type of particle plays an important role in the primordial universe.

Note that, although we fix $k_0=k_{\rm pivot}$ in
Tables~\ref{fig:resultBB_gm} and \ref{fig:resultBB_gp}, the
uncertainties with 
different $k_0$ can be easily obtained by use of the scaling, $\sigma_{g_n},
\sigma_{q_{LM}} \propto k_0^\gamma$, since Eq.~(\ref{eq:Cl1l2}) is
proportional to $k_0^{-\gamma}$ and 
$\sigma_{g_n}, \sigma_{q_{LM}}$ is thus proportional to the inverse of Eq.~(\ref{eq:Cl1l2}).

%%%%%%%%%%%%%%%%%%%%%%%%%%%%%%%%%%%%%%%%%%%%%%%%%%%%%%%%%%%%%%%%%%%%%%%%%%%%%%%
\section{Conclusion}
\label{sec:conclusion}
%%%%%%%%%%%%%%%%%%%%%%%%%%%%%%%%%%%%%%%%%%%%%%%%%%%%%%%%%%%%%%%%%%%%%%%%%%%%%%%
%%%%%%%%%%%%%%%%%%%%%%%%%%%%%%%%%%%%%%%%%%%%%%%%%%%%%%%%%%%%%%%%%%%%%%%%%%%%%%%

We investigated the detectability of the statistical anisotropies of the primordial tensor 
power spectrum using the Fisher information matrix assuming the
observations by CMB-S4, LiteBIRD and Planck. We parameterize the
primordial tensor power spectrum in Eq.~(\ref{eq:aniso1}) with
Eq.~(\ref{eq:sdep}) and Eq.(\ref{eq:aniso2}),
and estimate the 1$\sigma$-uncertainties of $q_{LM}$ and $g_n$ given in
Eq.~(\ref{eq:def_error}) with the fiducial values $q_{LM}=g_n=0$ for $L$
(or $n$) $\geq 2$ in the case of $\gamma=0$, and $q_{LM}=g_n=0$ for $L$
(or $n$) $\geq 0$ in the case of $\gamma\ne0$.

Our results are tabulated in Tables~\ref{fig:resultBB_g0}-\ref{fig:resultBB_gp}.
In Table \ref{fig:resultBB_g0}, we find that a relatively large
statistical anisotropy $g_n\sim\mathcal{O}(0.1)$ would possibly be detected by
CMB-S4 as long as we take into account up to $g_4$ and 
by LiteBIRD up to $g_2$, whereas
unfortunately the results imply difficulties to detect anisotropies by
Planck since it is contaminated by large noises. In addition, in order
to detect a higher multipole coefficient $g_6$, we need 
further observatories whose noise level is much more suppressed than CMB-S4.

Note that, in our present study, we do not take the contamination from
the CMB-lensing, unwanted B-mode signal converted from E-mode through the
gravitational interaction, into account. In the actual observations, the
detectability of the anisotropies depends on how well we can remove the
contamination, namely, delensing. Roughly speaking, this effect can be
included in our result by increasing $w_{BB}^{-1/2}$ defined in Eq.~(\ref{eq:def_NBB}).
If the delensing is not perfectly performed, the detectability of the
anisotropies by CMB-S4 would be worse.

\begin{acknowledgements}
 This work was partially supported by JSPS KAKENHI No. JP16K17695 (T.H.), JP15K17659 (S.Y.) and JP17J09103 (T.F.),  and also MEXT KAKENHI No. 15H05888 (S.Y.) and 18H04356 (S.Y.).
 T.H. was also supported by MEXT Supported Program for the Strategic Research Foundation at  Private Universities, 2014-2018 (S1411024).
\end{acknowledgements}

\appendix

%%%%%%%%%%%%%%%%%%%%%%%%%%%%%%%%%%%%%%%%%%%%%%%%%%%%%%%%%%%%%%%%%%%%%%%%%%%%%%%
%%%%%%%%%%%%%%%%%%%%%%%%%%%%%%%%%%%%%%%%%%%%%%%%%%%%%%%%%%%%%%%%%%%%%%%%%%%%%%%
\section{Fisher information matrix in the anisotropic cases}
\label{app:fisher}
%%%%%%%%%%%%%%%%%%%%%%%%%%%%%%%%%%%%%%%%%%%%%%%%%%%%%%%%%%%%%%%%%%%%%%%%%%%%%%%
%%%%%%%%%%%%%%%%%%%%%%%%%%%%%%%%%%%%%%%%%%%%%%%%%%%%%%%%%%%%%%%%%%%%%%%%%%%%%%%

The covariance matrix taking into account the correlations between
different $\ell$'s is given by
%%%%%%%%%%%%%%%%%%%%%%%%%%%%%%%%%%%%%%%%%%%%%%%%%%%%%%%%%%%%%%%%%%%%
%
\begin{align}
 \boldsymbol{C}_{\ell_1m_1;\ell_2m_2} = \begin{pmatrix}
        C_{\ell_1m_1;\ell_2m_2}^{TT} & C_{\ell_1m_1;\ell_2m_2}^{TE} & 0 \\
        C_{\ell_1m_1;\ell_2m_2}^{TE} & C_{\ell_1m_1;\ell_2m_2}^{EE} & 0 \\
        0 & 0 & C_{\ell_1m_1;\ell_2m_2}^{BB}
     \end{pmatrix}.
\end{align}
%
%%%%%%%%%%%%%%%%%%%%%%%%%%%%%%%%%%%%%%%%%%%%%%%%%%%%%%%%%%%%%%%%%%%%
The Fisher information matrix based on this covariance matrix is given as \cite{Verde:2009tu,Ma:2011ii}
%%%%%%%%%%%%%%%%%%%%%%%%%%%%%%%%%%%%%%%%%%%%%%%%%%%%%%%%%%%%%%%%%%%%
%
\begin{align}
 F_{ij} &= \frac{f_{\rm sky}}{2}\sum_{\ell_1m_1}\sum_{\ell_2m_2} {\rm Tr}\left[
    \boldsymbol{C}_{\ell_1}^{-1}
    \frac{\partial \boldsymbol{C}_{\ell_1m_1;\ell_2m_2}}{\partial\theta_i}
    \boldsymbol{C}_{\ell_2}^{-1}
    \frac{\partial \boldsymbol{C}_{\ell_2m_2;\ell_1m_1}}{\partial\theta_j}\right] \\
&= f_{\rm sky}\sum_{XY}\sum_{\ell_1m_1}\sum_{\ell_2m_2}
    \frac{\partial C_{\ell_1m_1;\ell_2m_2}^X}{\partial\theta_i}
    (\mathscr{C}_{\ell_1\ell_2}^{-1})^{XY}
    \frac{\partial C_{\ell_2m_2;\ell_1m_1}^Y}{\partial\theta_j},
\end{align}
%
%%%%%%%%%%%%%%%%%%%%%%%%%%%%%%%%%%%%%%%%%%%%%%%%%%%%%%%%%%%%%%%%%%%%
where $f_{\rm sky}$ denotes the fraction of the sky covered,
$X,Y=TT,TE,EE,BB$ and 
\begin{widetext}
%%%%%%%%%%%%%%%%%%%%%%%%%%%%%%%%%%%%%%%%%%%%%%%%%%%%%%%%%%%%%%%%%%%%
%
\begin{align}
\mathscr{C}_{\ell_1\ell_2}^{-1}
= \frac{1}{2}
\begin{pmatrix}
  C^{EE}_{\ell_1}C^{EE}_{\ell_2}/\Delta_{\ell_1\ell_2}
    & -\mathcal{C}^{TE,EE}_{\ell_1\ell_2}/\Delta_{\ell_1\ell_2}
    & C^{TE}_{\ell_1}C^{TE}_{\ell_2}/\Delta_{\ell_1\ell_2}
    & 0 \\
  -\mathcal{C}^{TE,EE}_{\ell_1\ell_2}/\Delta_{\ell_1\ell_2}
    & (\mathcal{C}^{TT,EE}_{\ell_1\ell_2}+2C^{TE}_{\ell_1}C^{TE}_{\ell_2})/\Delta_{\ell_1\ell_2}
    & -\mathcal{C}^{TT,TE}_{\ell_1\ell_2}/\Delta_{\ell_1\ell_2}
    & 0 \\
  C^{TE}_{\ell_1}C^{TE}_{\ell_2}/\Delta_{\ell_1\ell_2}
    & -\mathcal{C}^{TT,TE}_{\ell_1\ell_2}/\Delta_{\ell_1\ell_2}
    & C^{TT}_{\ell_1}C^{TT}_{\ell_2}/\Delta_{\ell_1\ell_2}
    & 0 \\
  0 & 0 & 0 & \frac{1}{C^{BB}_{\ell_1}C^{BB}_{\ell_2}}
\end{pmatrix},
 \label{eq:inv_cov_aniso}
\end{align}
%
%%%%%%%%%%%%%%%%%%%%%%%%%%%%%%%%%%%%%%%%%%%%%%%%%%%%%%%%%%%%%%%%%%%%
\end{widetext}
where 
%%%%%%%%%%%%%%%%%%%%%%%%%%%%%%%%%%%%%%%%%%%%%%%%%%%%%%%%%%%%%%%%%%%%
%
\begin{align}
 \Delta_{\ell_1\ell_2} &= 
\left[C^{TT}_{\ell_1}C^{EE}_{\ell_1}-\left(C^{TE}_{\ell_1}\right)^2\right]\left[C^{TT}_{\ell_2}C^{EE}_{\ell_2}-\left(C^{TE}_{\ell_2}\right)^2\right],\\
 \mathcal{C}^{XY}_{\ell_1\ell_2} &= C^{X}_{\ell_1}C^{Y}_{\ell_2}+C^{Y}_{\ell_1}C^{X}_{\ell_2},
\end{align}
%
%%%%%%%%%%%%%%%%%%%%%%%%%%%%%%%%%%%%%%%%%%%%%%%%%%%%%%%%%%%%%%%%%%%%
and
%%%%%%%%%%%%%%%%%%%%%%%%%%%%%%%%%%%%%%%%%%%%%%%%%%%%%%%%%%%%%%%%%%%%
%
\begin{align}
\widetilde{C}^{X}_{\ell} := C^{X}_{\ell} + \mathcal{N}^{X}_{\ell},
\label{eq:totalCl}
\end{align}
%
%%%%%%%%%%%%%%%%%%%%%%%%%%%%%%%%%%%%%%%%%%%%%%%%%%%%%%%%%%%%%%%%%%%%
with $\mathcal{N}^{X}_{\ell}$ being the noises for the detection of $X=TT,TE,EE,BB$.
The angular power spectrum $C^X_\ell$ with $X=BB$ 
is given by Eq.~(\ref{eq:Clmlm1}) with $\ell_1=\ell_2=\ell$ and $\gamma=0$,
and those of $X=TT,TE,EE$ can be also calculated by replacing the transfer 
functions in Eq.~(\ref{eq:Clmlm1}) with the corresponding ones.
If we choose $\{\theta_i\}=\{q_{LM}\}$, the Fisher matrix becomes
%%%%%%%%%%%%%%%%%%%%%%%%%%%%%%%%%%%%%%%%%%%%%%%%%%%%%%%%%%%%%%%%%%%%
%
\begin{align}
 F_{LM;L'M'} 
&= \delta_{LL'}\delta_{MM'}F_L,
\label{eq:Fisher_aniso_q}
\end{align}
%
%%%%%%%%%%%%%%%%%%%%%%%%%%%%%%%%%%%%%%%%%%%%%%%%%%%%%%%%%%%%%%%%%%%%
where
%%%%%%%%%%%%%%%%%%%%%%%%%%%%%%%%%%%%%%%%%%%%%%%%%%%%%%%%%%%%%%%%%%%%
%
\begin{align}
F_L &= \frac{f_{\rm sky}}{4\pi} \sum_{s_1s_2s_3s_4}
    (-1)^{s_1+s_3}
    \sum_{\ell_1\ell_2}
(2l_1+1)(2l_2+1)
\notag \\ &\times
\begin{pmatrix}
 l_1 & l_2 & L \\ 
 -s_1 & s_2 & s_1-s_2 
\end{pmatrix}
\begin{pmatrix}
 l_1 & l_2 & L \\ 
 -s_3 & s_4 & s_3-s_4 
\end{pmatrix}
\notag \\ &\quad\times
    \sum_{XY}C^{s_1s_2X}_{\ell_1\ell_2}(\mathscr{C}_{\ell_1\ell_2}^{-1})^{XY}C^{s_3s_4Y}_{\ell_2\ell_1},
\label{eq:Fisher_kernel}
\end{align}
%
%%%%%%%%%%%%%%%%%%%%%%%%%%%%%%%%%%%%%%%%%%%%%%%%%%%%%%%%%%%%%%%%%%%%
Alternatively, if we choose $\{\theta_i\}=\{g_{n}\}$, we have
%%%%%%%%%%%%%%%%%%%%%%%%%%%%%%%%%%%%%%%%%%%%%%%%%%%%%%%%%%%%%%%%%%%%
%
\begin{align}
 F_{mn} 
&= \sum_{LM}\frac{\partial q_{LM}}{\partial g_m}\frac{\partial q_{LM}}{\partial g_n}
   F^{XY}_{L}.
\label{eq:Fisher_aniso_g}
\end{align}
%
%%%%%%%%%%%%%%%%%%%%%%%%%%%%%%%%%%%%%%%%%%%%%%%%%%%%%%%%%%%%%%%%%%%%
The coefficients in the right-hand side are found to be 
%%%%%%%%%%%%%%%%%%%%%%%%%%%%%%%%%%%%%%%%%%%%%%%%%%%%%%%%%%%%%%%%%%%%
%
\begin{align}
 \frac{\partial q_{LM}}{\partial g_n} 
    = \delta_{M0}\sqrt{2L+1}\sqrt{\pi}\int^1_{-1}\!\mu^nP_L(\mu)\,d\mu,
\label{eq:qg}
\end{align}
%
%%%%%%%%%%%%%%%%%%%%%%%%%%%%%%%%%%%%%%%%%%%%%%%%%%%%%%%%%%%%%%%%%%%%
where $P_L(\mu)$ is the Legendre polynomials.


\begin{thebibliography}{5}

%\cite{Ade:2015lrj}
\bibitem{Ade:2015lrj} 
  P.~A.~R.~Ade {\it et al.} [Planck Collaboration],
  %``Planck 2015 results. XX. Constraints on inflation,''
  Astron.\ Astrophys.\  {\bf 594}, A20 (2016)
  doi:10.1051/0004-6361/201525898
  [arXiv:1502.02114 [astro-ph.CO]].
  %%CITATION = doi:10.1051/0004-6361/201525898;%%
  %1716 citations counted in INSPIRE as of 30 Jul 2018


%\cite{Array:2015xqh}
\bibitem{Array:2015xqh} 
  P.~A.~R.~Ade {\it et al.} [BICEP2 and Keck Array Collaborations],
  %``Improved Constraints on Cosmology and Foregrounds from BICEP2 and Keck Array Cosmic Microwave Background Data with Inclusion of 95 GHz Band,''
  Phys.\ Rev.\ Lett.\  {\bf 116}, 031302 (2016)
  doi:10.1103/PhysRevLett.116.031302
  [arXiv:1510.09217 [astro-ph.CO]].
  %%CITATION = doi:10.1103/PhysRevLett.116.031302;%%
  %425 citations counted in INSPIRE as of 30 Jul 2018

%\cite{Akrami:2018odb}
\bibitem{Akrami:2018odb} 
  Y.~Akrami {\it et al.} [Planck Collaboration],
  %``Planck 2018 results. X. Constraints on inflation,''
  arXiv:1807.06211 [astro-ph.CO].
  %%CITATION = ARXIV:1807.06211;%%
  %9 citations counted in INSPIRE as of 07 Aug 2018


%\cite{Matsumura:2013aja}
\bibitem{Matsumura:2013aja} 
  T.~Matsumura {\it et al.},
  %``Mission design of LiteBIRD,''
  J.\ Low.\ Temp.\ Phys.\  {\bf 176}, 733 (2014)
  doi:10.1007/s10909-013-0996-1
  [arXiv:1311.2847 [astro-ph.IM]].
  %%CITATION = doi:10.1007/s10909-013-0996-1;%%
  %172 citations counted in INSPIRE as of 30 Jul 2018


%\cite{Abazajian:2016yjj}
\bibitem{Abazajian:2016yjj} 
  K.~N.~Abazajian {\it et al.} [CMB-S4 Collaboration],
  %``CMB-S4 Science Book, First Edition,''
  arXiv:1610.02743 [astro-ph.CO].
  %%CITATION = ARXIV:1610.02743;%%
  %306 citations counted in INSPIRE as of 30 Jul 2018


%\cite{Namba:2015gja}
\bibitem{Namba:2015gja} 
  R.~Namba, M.~Peloso, M.~Shiraishi, L.~Sorbo and C.~Unal,
  %``Scale-dependent gravitational waves from a rolling axion,''
  JCAP {\bf 1601}, no. 01, 041 (2016)
  doi:10.1088/1475-7516/2016/01/041
  [arXiv:1509.07521 [astro-ph.CO]].
  %%CITATION = doi:10.1088/1475-7516/2016/01/041;%%
  %55 citations counted in INSPIRE as of 30 Jul 2018


%\cite{Dimastrogiovanni:2016fuu}
\bibitem{Dimastrogiovanni:2016fuu} 
  E.~Dimastrogiovanni, M.~Fasiello and T.~Fujita,
  %``Primordial Gravitational Waves from Axion-Gauge Fields Dynamics,''
  JCAP {\bf 1701}, no. 01, 019 (2017)
  doi:10.1088/1475-7516/2017/01/019
  [arXiv:1608.04216 [astro-ph.CO]].
  %%CITATION = doi:10.1088/1475-7516/2017/01/019;%%
  %31 citations counted in INSPIRE as of 30 Jul 2018


\bibitem{Shiraishi:2016yun} 
  M.~Shiraishi, C.~Hikage, R.~Namba, T.~Namikawa and M.~Hazumi,
  %``Testing statistics of the CMB B -mode polarization toward unambiguously establishing quantum fluctuation of the vacuum,''
  Phys.\ Rev.\ D {\bf 94}, no. 4, 043506 (2016)
  doi:10.1103/PhysRevD.94.043506
  [arXiv:1606.06082 [astro-ph.CO]].
  %%CITATION = doi:10.1103/PhysRevD.94.043506;%%
  %11 citations counted in INSPIRE as of 08 Aug 2018

%\cite{Agrawal:2017awz}
\bibitem{Agrawal:2017awz} 
  A.~Agrawal, T.~Fujita and E.~Komatsu,
  %``Large tensor non-Gaussianity from axion-gauge field dynamics,''
  Phys.\ Rev.\ D {\bf 97}, no. 10, 103526 (2018)
  doi:10.1103/PhysRevD.97.103526
  [arXiv:1707.03023 [astro-ph.CO]].

%\cite{Watanabe:2009ct}
\bibitem{Watanabe:2009ct} 
  M.~a.~Watanabe, S.~Kanno and J.~Soda,
  %``Inflationary Universe with Anisotropic Hair,''
  Phys.\ Rev.\ Lett.\  {\bf 102}, 191302 (2009)
  doi:10.1103/PhysRevLett.102.191302
  [arXiv:0902.2833 [hep-th]].
  %%CITATION = doi:10.1103/PhysRevLett.102.191302;%%
  %249 citations counted in INSPIRE as of 30 Jul 2018


%\cite{Watanabe:2010fh}
\bibitem{Watanabe:2010fh} 
  M.~a.~Watanabe, S.~Kanno and J.~Soda,
  %``The Nature of Primordial Fluctuations from Anisotropic Inflation,''
  Prog.\ Theor.\ Phys.\  {\bf 123}, 1041 (2010)
  doi:10.1143/PTP.123.1041
  [arXiv:1003.0056 [astro-ph.CO]].
  %%CITATION = doi:10.1143/PTP.123.1041;%%
  %141 citations counted in INSPIRE as of 30 Jul 2018


%\cite{Bartolo:2012sd}
\bibitem{Bartolo:2012sd} 
  N.~Bartolo, S.~Matarrese, M.~Peloso and A.~Ricciardone,
  %``Anisotropic power spectrum and bispectrum in the $f(\phi)F^2$ mechanism,''
  Phys.\ Rev.\ D {\bf 87}, no. 2, 023504 (2013)
  doi:10.1103/PhysRevD.87.023504
  [arXiv:1210.3257 [astro-ph.CO]].
  %%CITATION = doi:10.1103/PhysRevD.87.023504;%%
  %131 citations counted in INSPIRE as of 30 Jul 2018


%\cite{Fujita:2017lfu}
\bibitem{Fujita:2017lfu} 
  T.~Fujita and I.~Obata,
  %``Does anisotropic inflation produce a small statistical anisotropy?,''
  JCAP {\bf 1801}, no. 01, 049 (2018)
  doi:10.1088/1475-7516/2018/01/049
  [arXiv:1711.11539 [astro-ph.CO]].
  %%CITATION = doi:10.1088/1475-7516/2018/01/049;%%
  %5 citations counted in INSPIRE as of 30 Jul 2018


%\cite{Abolhasani:2013bpa}
\bibitem{Abolhasani:2013bpa} 
  A.~A.~Abolhasani, R.~Emami and H.~Firouzjahi,
  %``Primordial Anisotropies in Gauged Hybrid Inflation,''
  JCAP {\bf 1405}, 016 (2014)
  doi:10.1088/1475-7516/2014/05/016
  [arXiv:1311.0493 [hep-th]].
  %%CITATION = doi:10.1088/1475-7516/2014/05/016;%%
  %17 citations counted in INSPIRE as of 30 Jul 2018


%\cite{Bartolo:2013msa}
\bibitem{Bartolo:2013msa} 
  N.~Bartolo, S.~Matarrese, M.~Peloso and A.~Ricciardone,
  %``Anisotropy in solid inflation,''
  JCAP {\bf 1308}, 022 (2013)
  doi:10.1088/1475-7516/2013/08/022
  [arXiv:1306.4160 [astro-ph.CO]].
  %%CITATION = doi:10.1088/1475-7516/2013/08/022;%%
  %69 citations counted in INSPIRE as of 30 Jul 2018


%\cite{Akhshik:2014gja}
\bibitem{Akhshik:2014gja} 
  M.~Akhshik, R.~Emami, H.~Firouzjahi and Y.~Wang,
  %``Statistical Anisotropies in Gravitational Waves in Solid Inflation,''
  JCAP {\bf 1409}, 012 (2014)
  doi:10.1088/1475-7516/2014/09/012
  [arXiv:1405.4179 [astro-ph.CO]].
  %%CITATION = doi:10.1088/1475-7516/2014/09/012;%%
  %25 citations counted in INSPIRE as of 30 Jul 2018


%\cite{Bartolo:2014xfa}
\bibitem{Bartolo:2014xfa} 
  N.~Bartolo, M.~Peloso, A.~Ricciardone and C.~Unal,
  %``The expected anisotropy in solid inflation,''
  JCAP {\bf 1411}, no. 11, 009 (2014)
  doi:10.1088/1475-7516/2014/11/009
  [arXiv:1407.8053 [astro-ph.CO]].
  %%CITATION = doi:10.1088/1475-7516/2014/11/009;%%
  %24 citations counted in INSPIRE as of 30 Jul 2018


%\cite{Kehagias:2017cym}
\bibitem{Kehagias:2017cym} 
  A.~Kehagias and A.~Riotto,
  %``On the Inflationary Perturbations of Massive Higher-Spin Fields,''
  JCAP {\bf 1707}, no. 07, 046 (2017)
  doi:10.1088/1475-7516/2017/07/046
  [arXiv:1705.05834 [hep-th]].
  %%CITATION = doi:10.1088/1475-7516/2017/07/046;%%
  %17 citations counted in INSPIRE as of 30 Jul 2018


%\cite{Franciolini:2017ktv}
\bibitem{Franciolini:2017ktv} 
  G.~Franciolini, A.~Kehagias and A.~Riotto,
  %``Imprints of Spinning Particles on Primordial Cosmological Perturbations,''
  JCAP {\bf 1802}, no. 02, 023 (2018)
  doi:10.1088/1475-7516/2018/02/023
  [arXiv:1712.06626 [hep-th]].
  %%CITATION = doi:10.1088/1475-7516/2018/02/023;%%
  %13 citations counted in INSPIRE as of 30 Jul 2018


%\cite{Watanabe:2010bu}
\bibitem{Watanabe:2010bu} 
  M.~a.~Watanabe, S.~Kanno and J.~Soda,
  %``Imprints of Anisotropic Inflation on the Cosmic Microwave Background,''
  Mon.\ Not.\ Roy.\ Astron.\ Soc.\  {\bf 412}, L83 (2011)
  doi:10.1111/j.1745-3933.2011.01010.x
  [arXiv:1011.3604 [astro-ph.CO]].
  %%CITATION = doi:10.1111/j.1745-3933.2011.01010.x;%%
  %77 citations counted in INSPIRE as of 30 Jul 2018


%\cite{Chen:2014eua}
\bibitem{Chen:2014eua} 
  X.~Chen, R.~Emami, H.~Firouzjahi and Y.~Wang,
  %``The TT, TB, EB and BB correlations in anisotropic inflation,''
  JCAP {\bf 1408}, 027 (2014)
  doi:10.1088/1475-7516/2014/08/027
  [arXiv:1404.4083 [astro-ph.CO]].
  %%CITATION = doi:10.1088/1475-7516/2014/08/027;%%
  %38 citations counted in INSPIRE as of 30 Jul 2018


%\cite{Bartolo:2014hwa}
\bibitem{Bartolo:2014hwa} 
  N.~Bartolo, S.~Matarrese, M.~Peloso and M.~Shiraishi,
  %``Parity-violating and anisotropic correlations in pseudoscalar inflation,''
  JCAP {\bf 1501}, no. 01, 027 (2015)
  doi:10.1088/1475-7516/2015/01/027
  [arXiv:1411.2521 [astro-ph.CO]].
  %%CITATION = doi:10.1088/1475-7516/2015/01/027;%%
  %31 citations counted in INSPIRE as of 30 Jul 2018


%\cite{Bartolo:2017sbu}
\bibitem{Bartolo:2017sbu} 
  N.~Bartolo, A.~Kehagias, M.~Liguori, A.~Riotto, M.~Shiraishi and V.~Tansella,
  %``Detecting higher spin fields through statistical anisotropy in the CMB and galaxy power spectra,''
  Phys.\ Rev.\ D {\bf 97}, no. 2, 023503 (2018)
  doi:10.1103/PhysRevD.97.023503
  [arXiv:1709.05695 [astro-ph.CO]].
  %%CITATION = doi:10.1103/PhysRevD.97.023503;%%
  %12 citations counted in INSPIRE as of 30 Jul 2018


%\cite{Franciolini:2018eno}
\bibitem{Franciolini:2018eno} 
  G.~Franciolini, A.~Kehagias, A.~Riotto and M.~Shiraishi,
  %``Detecting higher spin fields through statistical anisotropy in the CMB bispectrum,''
  arXiv:1803.03814 [astro-ph.CO].
  %%CITATION = ARXIV:1803.03814;%%
  %5 citations counted in INSPIRE as of 30 Jul 2018


%\cite{Kim:2013gka}
\bibitem{Kim:2013gka} 
  J.~Kim and E.~Komatsu,
  %``Limits on anisotropic inflation from the Planck data,''
  Phys.\ Rev.\ D {\bf 88}, 101301 (2013)
  doi:10.1103/PhysRevD.88.101301
  [arXiv:1310.1605 [astro-ph.CO]].
  %%CITATION = doi:10.1103/PhysRevD.88.101301;%%
  %86 citations counted in INSPIRE as of 30 Jul 2018


%\cite{Fujita:2018zbr}
\bibitem{Fujita:2018zbr} 
  T.~Fujita, I.~Obata, T.~Tanaka and S.~Yokoyama,
  %``Statistically Anisotropic Tensor Modes from Inflation,''
  JCAP {\bf 1807}, no. 07, 023 (2018)
  doi:10.1088/1475-7516/2018/07/023
  [arXiv:1801.02778 [astro-ph.CO]].
  %%CITATION = doi:10.1088/1475-7516/2018/07/023;%%

%\cite{Obata:2018ilf}
\bibitem{Obata:2018ilf} 
  I.~Obata and T.~Fujita,
  %``Footprint of Two-Form Field: Statistical Anisotropy in Primordial Gravitational Waves,''
  arXiv:1808.00548 [astro-ph.CO].
  %%CITATION = ARXIV:1808.00548;%%


%\cite{Shiraishi:2014owa}
\bibitem{Shiraishi:2014owa} 
  M.~Shiraishi, D.~F.~Mota, A.~Ricciardone and F.~Arroja,
  %``CMB statistical anisotropy from noncommutative gravitational waves,''
  JCAP {\bf 1407}, 047 (2014)
  doi:10.1088/1475-7516/2014/07/047
  [arXiv:1401.7936 [astro-ph.CO]].
  %%CITATION = doi:10.1088/1475-7516/2014/07/047;%%
  %6 citations counted in INSPIRE as of 30 Jul 2018


%\cite{Bartolo:2018igk}
\bibitem{Bartolo:2018igk} 
  N.~Bartolo, A.~Hoseinpour, G.~Orlando, S.~Matarrese and M.~Zarei,
  %``Photon-graviton scattering: A new way to detect anisotropic gravitational waves?,''
  Phys.\ Rev.\ D {\bf 98}, no. 2, 023518 (2018)
  doi:10.1103/PhysRevD.98.023518
  [arXiv:1804.06298 [gr-qc]].
  %%CITATION = doi:10.1103/PhysRevD.98.023518;%%


%\cite{Valentini:2015sna}
\bibitem{Valentini:2015sna} 
  A.~Valentini,
  %``Statistical anisotropy and cosmological quantum relaxation,''
  arXiv:1510.02523 [astro-ph.CO].
  %%CITATION = ARXIV:1510.02523;%%
  %7 citations counted in INSPIRE as of 30 Jul 2018


%\cite{Katayama:2011eh}
\bibitem{Katayama:2011eh} 
  N.~Katayama and E.~Komatsu,
  %``Simple foreground cleaning algorithm for detecting primordial B-mode polarization of the cosmic microwave background,''
  Astrophys.\ J.\  {\bf 737}, 78 (2011)
  doi:10.1088/0004-637X/737/2/78
  [arXiv:1101.5210 [astro-ph.CO]].
  %%CITATION = doi:10.1088/0004-637X/737/2/78;%%
  %36 citations counted in INSPIRE as of 30 Jul 2018


%\cite{Hiramatsu:2018nfa}
\bibitem{Hiramatsu:2018nfa} 
  T.~Hiramatsu, E.~Komatsu, M.~Hazumi and M.~Sasaki,
  %``Reconstruction of primordial tensor power spectra from B-mode polarization of the cosmic microwave background,''
  Phys.\ Rev.\ D {\bf 97}, no. 12, 123511 (2018)
  doi:10.1103/PhysRevD.97.123511
  [arXiv:1803.00176 [astro-ph.CO]].
  %%CITATION = doi:10.1103/PhysRevD.97.123511;%%


%\cite{Ade:2015xua}
\bibitem{Ade:2015xua} 
  P.~A.~R.~Ade {\it et al.} [Planck Collaboration],
  %``Planck 2015 results. XIII. Cosmological parameters,''
  Astron.\ Astrophys.\  {\bf 594}, A13 (2016)
  doi:10.1051/0004-6361/201525830
  [arXiv:1502.01589 [astro-ph.CO]].
  %%CITATION = doi:10.1051/0004-6361/201525830;%%
  %5939 citations counted in INSPIRE as of 30 Jul 2018


%\cite{Planck:2006aa}
\bibitem{Planck:2006aa} 
  J.~Tauber {\it et al.} [Planck Collaboration],
  %``The Scientific programme of Planck,''
  astro-ph/0604069.
  %%CITATION = ASTRO-PH/0604069;%%
  %630 citations counted in INSPIRE as of 30 Jul 2018


%\cite{Verde:2009tu}
\bibitem{Verde:2009tu}
  L.~Verde,
  %``Statistical methods in cosmology,''
  Lect.\ Notes Phys.\  {\bf 800} (2010) 147
  %doi:10.1007/978-3-642-10598-2_4
  [arXiv:0911.3105 [astro-ph.CO]].
  %%CITATION = doi:10.1007/978-3-642-10598-2_4;%%
  %41 citations counted in INSPIRE as of 15 Aug 2018

%\cite{Ma:2011ii}
\bibitem{Ma:2011ii} 
  Y.~Z.~Ma, G.~Efstathiou and A.~Challinor,
  %``Testing a direction-dependent primordial power spectrum with observations of the Cosmic Microwave Background,''
  Phys.\ Rev.\ D {\bf 83}, no. 8, 083005 (2011)
  Erratum: [Phys.\ Rev.\ D {\bf 89}, no. 12, 129901 (2014)]
  doi:10.1103/PhysRevD.89.129901, 10.1103/PhysRevD.83.083005
  [arXiv:1102.4961 [astro-ph.CO]].
  %%CITATION = doi:10.1103/PhysRevD.89.129901, 10.1103/PhysRevD.83.083005;%%
  %36 citations counted in INSPIRE as of 30 Jul 2018

\end{thebibliography}
\end{document}